\documentclass[twocolumn,showpacs,preprintnumbers,amsmath,amssymb]{revtex4}

\usepackage{graphicx}
\usepackage{dcolumn}
\usepackage{bm}

\def\gsim{\mathop {\vtop {\ialign {##\crcr 
$\hfil \displaystyle {>}\hfil $\crcr \noalign {\kern1pt \nointerlineskip } 
$\,\sim$ \crcr \noalign {\kern1pt}}}}\limits}
\def\lsim{\mathop {\vtop {\ialign {##\crcr 
$\hfil \displaystyle {<}\hfil $\crcr \noalign {\kern1pt \nointerlineskip } 
$\,\,\sim$ \crcr \noalign {\kern1pt}}}}\limits}

\begin{document}

\preprint{APS/123-QED}

\title{Magnetic-Field Control of Quantum Critical Points of Valence Transition}

\author{Shinji Watanabe$^1$}
\author{Atsushi Tsuruta$^2$} 
\author{Kazumasa Miyake$^2$}
\author{Jacques Flouquet$^3$}
%
\affiliation{%
Department of Applied Physics, University of Tokyo, Hongo 7-3-1, Bunkyo-ku, Tokyo, 113-8656, Japan$^1$
\\
Division of Materials Physics, Department of Materials Engineering Science, Graduate School of
Engineering Science, Osaka University, Toyonaka, Osaka 560-8531, Japan$^2$
\\
D{\'e}partement de la Recherche Fondamentale sur la Mati{\`e}re Condense{\'e}, SPSMS, CEA Grenoble,
17 rue des Martyrs, 38054 Grenoble Cedex 9, France$^3$
}%

\date{\today}

\begin{abstract}
We study 
the mechanism how critical end points of first-order valence transitions are controlled 
by a magnetic field. 
We show that 
the critical temperature is suppressed to be a quantum critical point (QCP) by a magnetic field 
and unexpectedly the QCP exhibits nonmonotonic field dependence in the ground-state phase diagram, 
giving rise to emergence of metamagnetism even in the intermediate valence-crossover regime. 
The driving force of the field-induced QCP is clarified to be cooperative phenomena of 
Zeeman effect and Kondo effect, which create a distinct energy scale 
from the Kondo temperature. 
This mechanism explains peculiar magnetic response in $\rm CeIrIn_5$ and 
metamagnetic transition in $\rm YbXCu_4$ for X=In 
as well as sharp contrast between X=Ag and Cd. 
\end{abstract}

\maketitle

Quantum critical phenomena is one of the most interesting issues  
in condensed matter physics. 
Quantum critical points (QCP) emerging when continuous-transition temperature 
of symmetry breakings such as magnetic orders is suppressed to absolute zero 
have been studied intensively~\cite{Moriya,Hertz,Millis}, 
since the critical fluctuations induce unusual normal-state behaviors and 
even trigger other instabilities such as unconventional superconductivity~\cite{Stewart}. 

Valence transition is an isostructural phase transition, 
which is known to occur as $\gamma$-$\alpha$ transition in Ce metal~\cite{Cevalence} 
and also in $\rm YbInCu_4$~\cite{Felner} 
characterized by a jump of the valence of the Ce and Yb ion. 
At the critical end point of the first-order valence transition (FOVT),  
the valence fluctuation diverges as diverging density fluctuations 
in the liquid-gas transition. 
When the critical temperature is suppressed by tuning material parameters 
and enters the Fermi-degeneracy regime,  
diverging valence fluctuations are considered to be coupled with the Fermi-surface instability. 
This multiple instability seems to be a key mechanism 
which dominates the low-temperature properties of the materials 
including the valence-fluctuating ions such as Ce and Yb~\cite{OM,ML,Dzero,WIM}.  

Actually, a remarkable increase of the superconducting transition temperature 
far from antiferromagnetic QCP in $\rm CeCu_2X_2$ (X=Ge, Si)~\cite{jaccard,holms,yuan} 
and $\rm CeIrIn_5$~\cite{kawasaki} as well as linear-temperature dependence of 
resistivity observed in variety of Ce and Yb compounds~\cite{holms,yuan,kawasaki,YbInCu4, wada,M07} 
seems to a sign of underlying influence of the proximity of the QCP of the valence transition (VQCP)~\cite{M07}.

So far, to reduce the critical temperature of the valence transition 
toward absolute zero, intensive efforts have been made 
by chemical substitutions and applying pressure~\cite{fisk}. 
Usually, magnetic field also offers one of the efficient control parameters. 
However, it is highly nontrivial how the critical end point as well as the QCP is controlled by 
applying magnetic field, 
since valence instabilities are essentially ascribed to charge degrees of freedom, 
i.e., relative change of the 4f- and conduction-electron charges. 

In this Letter, we show that magnetic field is an efficient parameter to 
change the critical end point of the FOVT to the QCP. 
Unexpectedly, 
we discover non-monotonic field dependence of the VQCP in the ground-state 
phase diagram, whose mechanism is clarified to be cooperative phenomena 
by Zeeman effect and Kondo effect. 
We show that a metamagnetic jump in the magnetization 
is caused by the proximity effect of the VQCP by magnetic field 
even in the intermediate valence-crossover regime. 
This mechanism explains a peculiar magnetic response observed 
in $\rm CeIrIn_5$~\cite{takeuchi,Palm} 
as well as sharp contrast between $\rm YbAgCu_4$ and $\rm YbCdCu_4$~\cite{YbInCu4}. 
Our results indicate significance of a distinct energy scale, which is characterized by 
the closeness to the VQCP as a key parameter for the valence-fluctuating materials. 

\begin{figure}[h]
\includegraphics[width=50mm]{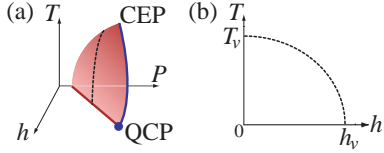}
\caption{\label{fig:TPH}(color online) (a) Schematic phase diagram, 
showing the FOVT surface in the $T$-$P$-$h$ space, where $P$ represents 
a control parameter (e.g. pressure, chemical concentration, etc.). 
The critical end point(CEP)s form a continuous transition line touched at $T=0$ as the QCP. 
(b) FOVT line $(h/h_{\rm v})^2+(T/T_{\rm v})^2=1$~\cite{magCe,Immer,Basu} 
in the $T$-$h$ plane for a fixed $P$, corresponding to the dashed line 
in (a). 
}
\end{figure}

First we demonstrate how the critical end point is suppressed to absolute zero. 
Let us consider the Claudius-Clapeyron 
relation for the FOVT temperature $T_{\rm v}$: 
$\delta T_{\rm v}/\delta h=-(m_{\rm K}-m_{\rm MV})/(S_{\rm K}-S_{\rm MV})$,  
where $m$ and $S$ denote the magnetization and the entropy, respectively. 
Here, K represents the Kondo regime where the f-electron (hole) density 
per site $\langle n_{\rm f} \rangle$ 
is close to 1 in the Ce (Yb) system, i.e., $\rm Ce^{3+} (4f^1)$ 
and $\rm Yb^{3+} (4f^{13})$, 
and MV represents the mixed-valence regime with $\langle n_{\rm f} \rangle <1$~\cite{defMV}. 
Since the magnetization as well as the entropy in the Kondo regime 
are larger than those in the MV regime, 
as observed in the specific heat and the uniform susceptibility, 
it turns out that 
$T_{\rm v}$ is suppressed by applying $h$ (see Fig.~\ref{fig:TPH}(a)). 
Then, the critical end point is suppressed to 
$T=0$
by $h$. 

Furthermore, the field dependence of $T_{\rm v}$ in the zero-temperature limit 
is also derived by using the above relation: 
For $T\to 0$, the entropy shows the $T$-linear behavior 
in both the Kondo and MV regimes 
and $S_{\rm K}-S_{\rm MV}$ is approximated to be proportional to $T_{\rm v}$ 
in case that $T_{\rm v}$ is smaller than the characteristic energy scales 
in the Kondo and MV regimes. 
Noting the temperature independent $m_{\rm K}-m_{\rm MV}$, we have 
$\delta T_{\rm v}/\delta h=-C_1/T_{\rm v}$, 
leading to $T_{\rm v}=C_2\sqrt{h_{\rm v}-h}$ with constants $C_1$ and $C_2$, 
which explains well the observed behavior in the Ce metal~\cite{magCe} 
and $\rm YbInCu_4$~\cite{Immer} (see Fig.~\ref{fig:TPH}(b)). 
We stress here that our analysis not only provides firm ground for small-$T_{\rm v}$ behavior
by considering the coherence of electrons which is essential for low temperature, 
but also interpolates high $T_{\rm v}$ 
satisfying the relation $(h/h_{\rm v})^2+(T/T_{\rm v})^2=1$~\cite{magCe,Immer,Basu} to zero temperature, 
since this relation was derived by assuming the isolated atomic entropy~\cite{magCe},
which is justified only at the high temperature regime. 

Although we have shown that the $h$ dependence of the critical end point can be understood 
from the viewpoint of the free-energy gain by the larger entropy in the Kondo regime, 
it is highly nontrivial how the VQCP is controlled by $h$ at $T=0$. 
To proceed our analysis, 
we introduce a minimal model which describes the essential part 
of Ce and Yb systems in the standard notation: 
\begin{equation}
H=H_{\rm c}+H_{\rm f}+H_{\rm hyb}+H_{U_{\rm fc}}-h\sum_{i}(S_i^{{\rm f}z}+S_i^{{\rm c}z}), 
\label{eq:PAM} 
\end{equation}
where 
$H_{\rm c}=\sum_{{\bf k}\sigma}\varepsilon_{\bf k}
c_{{\bf k}\sigma}^{\dagger}c_{{\bf k}\sigma}$, 
$H_{\rm f}=\varepsilon_{ \rm f}\sum_{i\sigma}n^{ \rm f}_{i\sigma}
+U\sum_{i=1}^{N}n_{i\uparrow}^{ \rm f}n_{i\downarrow}^{ \rm f}
$, 
$H_{\rm hyb}=V\sum_{i\sigma}\left(
f_{i\sigma}^{\dagger}c_{i\sigma}+c_{i\sigma}^{\dagger}f_{i\sigma}
\right)$ 
and 
$
H_{U_{\rm fc}}=
U_{\rm fc}\sum_{i=1}^{N}n_{i}^{ \rm f}n_{i}^{ c}
$. 
%
The $U_{\rm fc}$ term is the Coulomb repulsion between the f and conduction electrons, 
which is considered to play an important role in the valence transition~\cite{Falicov,hewson,singh,Zlatic,OM,WIM}. 
For example, in the case of Ce metal which exhibits the $\gamma$-$\alpha$ transition, 
the 4f- and 5d-electron bands are located at the Fermi level~\cite{Ceband}.   
Since both the orbitals are located on the same Ce site, this term cannot be neglected. 
In the case of $\rm YbInCu_4$, $H_{U_{\rm fc}}$ also plays a crucial role for the FOVT 
in the hole picture of eq.~(\ref{eq:PAM})~\cite{Zlatic}. 

We apply the slave-boson-mean-field theory~\cite{OM} 
to eq~(\ref{eq:PAM}). 
To describe the state for $U=\infty$, 
we consider 
$Vf_{i\sigma}^{\dagger}b_{i}c_{i\sigma}$ 
instead of 
$V f_{i\sigma}^{\dagger}c_{i\sigma}$ in eq.~(\ref{eq:PAM}) 
by introducing the slave-boson operator $b_{i}$ at the $i$-th site 
to describe the ${\rm f}^{0}$ state 
and require the constraint 
$
\sum_{i}\lambda_{i}\left(
\sum_{\sigma}n_{i\sigma}^{ \rm f}+b_{i}^{\dagger}b_{i}-1
\right)
$
with $\lambda_i$ being the Lagrange multiplier. 
For $H_{U_{\rm fc}}$ in eq.~(\ref{eq:PAM}), 
we employ the mean-field decoupling as 
%
$
n_{i}^{ \rm f}n_{i}^{ c}\simeq 
n_{ \rm f}n_{i}^{ c}+n_{ c}n_{i}^{ \rm f}-\frac{1}{2}n_{ \rm f}n_{ c}. 
$
%
By approximating mean fields as uniform ones, i.e.,  
$b=\langle b_{i} \rangle$ and $\bar{\lambda}=\lambda_i$, 
the set of mean-field equations 
is obtained by  
${\partial\langle H \rangle}/{\partial \lambda}=0$ and 
${\partial \langle H \rangle}/{\partial b}=0$: 
%
%
$
\bar{\lambda}
=
\frac{V^{2}}{N}
\sum_{{\bf k}\sigma}
\frac{f(E_{{\bf k}\sigma}^{-})-f(E_{{\bf k}\sigma}^{+})}
{\sqrt{(\bar{\varepsilon}_{{ \rm f}\sigma}
-\bar{\varepsilon}_{{\bf k}\sigma})^{2}+4{\bar{V}}^{2}}} 
$, 
$
1-
|\bar{b}|^{2}
=
\frac{1}{2N}
\sum_{{\bf k}\sigma,\pm}
\left[
1\pm\frac{\bar{\varepsilon}_{{ \rm f}\sigma}
-\bar{\varepsilon}_{{\bf k}\sigma}}
{\sqrt{(\bar{\varepsilon}_{{ \rm f}\sigma}
-\bar{\varepsilon}_{{\bf k}\sigma})^{2}+4\bar{V}^{2}}}
\right]
f(E^{\pm}_{{\bf k}\sigma})
$, 
and the following equation holds for the total electron number: 
$
\bar{n}_{ \rm f}+\bar{n}_{ c}
=
\sum_{{\bf k}\sigma}
\left[
f(E^{-}_{{\bf k}\sigma})+f(E^{+}_{{\bf k}\sigma})
\right]/N. 
$
%
Here, $f(E)$ is the Fermi distribution function and 
$E^{\pm}_{{\bf k}\sigma}$ are the lower $(-)$ and upper $(+)$ 
hybridized bands for the quasi particle with spin $\sigma$, respectively: 
%
$
E^{\pm}_{{\bf k}\sigma}=
\frac{1}{2}
\left[
\bar{\varepsilon}_{{ \rm f}\sigma}
+\bar{\varepsilon}_{{\bf k}\sigma}
\pm
\sqrt{(
\bar{\varepsilon}_{{ \rm f}\sigma}
-\bar{\varepsilon}_{{\bf k}\sigma})^{2}+4\bar{V}^{2}}
\right], 
$
%
where 
$\bar{\varepsilon}_{{\bf k}\sigma}$, $\bar{\varepsilon}_{{ \rm f}\sigma}$ and $\bar{V}$ 
are defined by 
$
\bar{\varepsilon}_{{\bf k}\sigma}\equiv
\varepsilon_{\bf k}
+U_{ \rm fc}\bar{n}_{ \rm f}
-\frac{h\sigma}{2}
$, 
$
\bar{\varepsilon}_{{ \rm f}\sigma}\equiv
\varepsilon_{ \rm f}
+\bar{\lambda}
+U_{ \rm fc}\bar{n}_{ c}
-\frac{h\sigma}{2}
$ 
and 
$
\bar{V}\equiv
V|\bar{b}|
$. 
The dispersion of the conduction electrons is taken as 
$\varepsilon_{\bf k}={\bf k}^2/(2m)-D$ with $D$ being the bottom of the conduction band 
and the density of states $N_{0}(\varepsilon)$ is set to satisfy the normalization 
condition, $\int_{-D}^{D}d\varepsilon N_{0}(\varepsilon)=1$ per spin 
in three dimension. 
We take $D$ as the energy unit and show the results for $V=0.5$ and the total filling 
$n=(\bar{n}_{\rm f}+\bar{n}_{{\rm c}})/2=7/8$. 

At $h=0$ and $T=0$, a jump in $\bar{n}_{ \rm f}$ appears as a function of $\varepsilon_{\rm f}$ 
for large $U_{\rm fc}$, 
which indicates the first-order quantum phase transition, 
since a jump in $\bar{n}_{ \rm f}$ results in the level crossing of the ground states 
by the relation 
$\bar{n}_{ \rm f}=\partial \langle H \rangle/\partial \varepsilon_{ \rm f}$. 
The FOVT 
between the Kondo state and the MV state is caused by $U_{ \rm fc}$, 
since a large $U_{ \rm fc}$ forces the electrons to pour into either the f level or the conduction band. 
The QCP in the $\varepsilon_{\rm f}$-$U_{\rm fc}$ plane is identified to be 
$(\varepsilon_{\rm f}^{\rm QCP}, U_{\rm fc}^{\rm QCP})=(0.356,1.464)$, 
at which the jump in $\bar{n}_{\rm f}$ disappears 
and the valence susceptibility 
$\chi_{\rm v}\equiv -\partial^{2} \langle H \rangle/\partial \varepsilon_{ \rm f}^{2}
=-\partial \bar{n}_{ \rm f}/\partial \varepsilon_{ \rm f}$ diverges. 
The characteristic energy scale of the system, the so-called Kondo temperature, 
which is defined as $T_{\rm K}\equiv \bar{\varepsilon}_{{\rm f} \sigma}-\mu$ 
is estimated to be $T_{\rm K}=0.074$ at the QCP. 

Note that the surface of the 
FOVT 
exists in the parameter 
space of $\varepsilon_{\rm f}$, $U_{\rm fc}$ and $V$ and a trajectory line is drawn 
in the space for corresponding experimental parameter such as pressure (i.e., $P$ in Fig.1). 
It is also noted that the valence instability is considered to be coupled to the 
phonon degrees of freedom. Hence, in our model (\ref{eq:PAM}) the effect of the hybridization 
also plays an important role, which might share common aspects with the 
Kondo-volume-collapse scenario for Ce metal~\cite{Dzero}. 

\begin{figure}[t]
\includegraphics[width=60mm]{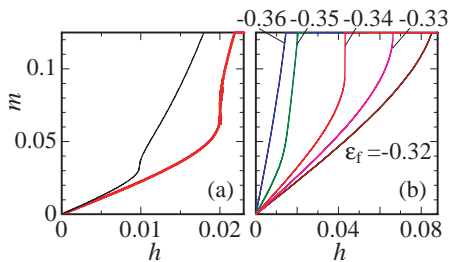} 
\caption{\label{fig:mhcurve}(color online)  
$m$-$h$ curve 
(a) for $(\varepsilon_{\rm f}, U_{\rm fc})=(-0.354,1.458)$ (thin line) 
and $(-0.349, 1.442)$ (bold line), 
and (b) for $\varepsilon_{\rm f}$ ranging from $-0.36$ to $-0.32$ 
at $U_{\rm fc}=1.42$. In both cases, $D=1$, $V=0.5$ at $n=7/8$. 
}
\end{figure}

\begin{figure}[t]
\includegraphics[width=61mm]{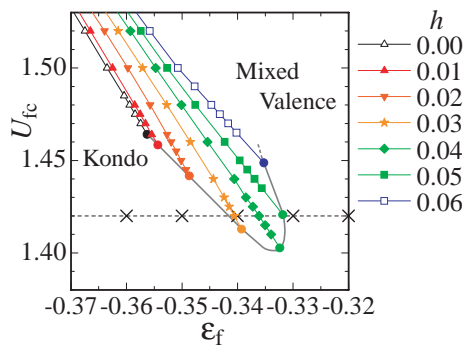}  
\caption{\label{fig:hPD}(color online) Ground-state phase diagram in the plane of $U_{\rm fc}$ and 
$\varepsilon_{\rm f}$ for $D=1$, $V=0.5$ at $n=7/8$. 
The FOVT line with a QCP for 
$h=0.00$ (open triangle), 0.01 (filled triangle) 0.02 (inverse triangle), 
0.03 (filled star), 0.04 (filled diamond), 0.05 (filled square) and 0.06 (open square). 
The shaded line connects the QCP under $h$, which is a guide for the eyes.
The dashed line with crosses represents $U_{\rm fc}=1.42$ (see text). 
}
\end{figure}

A remarkable result is found  under the magnetic field 
in the valence-crossover regime for $U_{\rm fc}<U_{\rm fc}^{\rm QCP}$. 
Figure~\ref{fig:mhcurve}(a) shows the magnetization 
$m=\sum_{i}\langle S_i^{{\rm f}z} + S_i^{{\rm c}z}\rangle/N$ vs. $h$ 
for $(\varepsilon_{\rm f}, U_{\rm fc})$$=$
$(-0.354,1.458)$
(thin line) 
and 
$(-0.349, 1.442)$ 
(bold line), indicating that 
the metamagnetism defined by the diverging magnetic susceptibility $\chi=\partial m/\partial h=\infty$ 
emerges at $h=h_{\rm m}=0.01$ and 0.02, respectively. 
To clarify the origin, we have determined the FOVT line as well as the QCP 
under the magnetic field. 
The result is shown in Fig.~\ref{fig:hPD}. 
It is found that the FOVT line extends to the MV regime and the location of the QCP is 
shifted to the smaller $U_{ \rm fc}$ and $|\varepsilon_{ \rm f}|$ direction, 
when $h$ is applied. 
This low-$h$ behavior of the FOVT line agrees with the low-temperature limit of $T_{\rm v}$ 
discussed above. 

We show the $m$-$h$ curve 
at $U_{\rm fc}=1.42$ for $\varepsilon_{\rm f}$ ranging from $-0.32$ to $-0.36$ 
in Fig.~\ref{fig:mhcurve}(b). 
$T_{\rm K}$ at $h=0$ is estimated as 
0.0353, 
0.0873, 
0.1346, 
0.1611, 
and 
0.1823 
for $\varepsilon_{\rm f}=-0.36$, $-0.35$, $-0.34$, $-0.33$ and $-0.32$, respectively  
(see crosses in Fig.~\ref{fig:hPD}). 
From these results, the mechanism is understood as follows:
At $h=0$, $T_{\rm K}$ is originally large for $\varepsilon_{\rm f}=-0.32$ and $-0.33$, 
since the system is in the MV regime. 
However, by applying $h$, the QCP approaches, which makes $T_{\rm K}$ reduced, 
since the system is forced to be closer to the Kondo regime by $h$. 
At the magnetic field $h=h_{\rm m}$ where the QCP is reached, the metamagnetism occurs 
with a singularity $\delta m \sim \delta h^{1/3}$~\cite{Millis2} 
as shown in Fig.~\ref{fig:mhcurve}(a).   
On the other hand, for $\varepsilon_{\rm f}=-0.35$ and $-0.36$, 
the QCP is not approached, so that no metamagnetism appears. 

An intriguing result is found in 
Fig.~\ref{fig:hPD}, which exhibits a non-monotonic $h$ dependence of the QCP:
As $h$ increases, 
the QCP shows an upturn around $h=0.04$, 
which is comparable to $T_{\rm K}$ at the QCP for $h=0$. 
The upturn of the QCP  
has been also confirmed for a constant density of states $N_{0}(\varepsilon)=1/(2D)$, 
suggesting that this behavior appears irrespective of details of the band structure.

The non-monotonic behavior can be understood from the structure of the valence 
susceptibility $\chi_{\rm v}$, which is given essentially by the RPA as discussed 
in ref.17.  Namely, it is given as 
$
\chi_{\rm v}(q)\approx {\chi_{ \rm fc}^{(0)}(q)
/[1-U_{ \rm fc}\chi_{ \rm fc}^{(0)}(q)}],
$
%
where $\chi_{ \rm fc}^{(0)}$ is the bubble diagram composed of f and 
conduction electrons.  In the Kondo regime ($h\lsim T_{ \rm K}$), where 
f electrons have a dominant spectral weight at around 
$\epsilon\sim-\varepsilon_{ \rm f}$ with 
width $\Delta\simeq \pi V^{2}N(\varepsilon_{ \rm F})$, $\chi_{ \rm fc}^{(0)}$ is 
estimated as $\chi_{ \rm fc}^{(0)}\approx 1/|\varepsilon_{ \rm f}|$ and is 
shown to be an increasing function of $h$.  Therefore, $U_{ \rm fc}^{\rm QCP}$ 
is decreasing as $h$ is applied until it reaches around $h\sim T_{ \rm K}$, and 
$|\varepsilon_{ \rm f}^{\rm QCP}|\approx U_{ \rm fc}^{\rm QCP}$ is also 
decreasing.  For $h\gsim T_{ \rm K}$, mass enhancement ($\sim 1/z$) is quickly 
suppressed and 
the MV regime 
is approached.  Then, 
$\chi_{ \rm fc}^{(0)}$ is given as 
$\chi_{ \rm fc}^{(0)}\approx 1/\Delta <1/|\varepsilon_{ \rm f}|$ with a help of 
shift of the f level towards the Fermi level, i.e., 
$\varepsilon_{ \rm f}\to\varepsilon_{ \rm f}+U_{ \rm fc}\delta
\langle n_{ c}\rangle$ ($\delta\langle n_{ c}\rangle$ being the 
change of number of conduction electrons per site due to entering 
the MV regime), 
so that $U_{ \rm fc}^{\rm QCP}$ is larger than that of 
$U_{ \rm fc}^{\rm QCP}(h\sim T_{ \rm K})$.  

We find that $h_{\rm m}$ corresponds to the difference of $T_{\rm K}$ at the QCP 
between $h=0$ and $h \ne 0$, i.e., 
$h_{\rm m}\sim \Delta T_{\rm K}^{\rm QCP}=
T_{\rm K}^{\rm QCP}(h \ne 0)-T_{\rm K}^{\rm QCP}(h=0)$. 
This indicates emergence of a new energy scale distinct from 
$T_{ \rm K}$, which is characterized by the closeness to the VQCP. 
Furthermore, 
the proximity of the VQCP to the intermediate valence-crossover regime under $h$ 
yields emergence of the metamagnetism as well as a jump in $m$ 
even without showing the temperature-driven FOVT at $h=0$.

This mechanism explains a peculiar behavior observed in $\rm CeIrIn_5$, 
which shows a jump in the $m$-$h$ curve at 42~T, but does not show 
the first-order transition in any physical quantities
when temperature is changed~\cite{takeuchi,Palm}. 
Namely, this is naturally understood 
if $\rm CeIrIn_5$ is located inside the enclosed area of the QCP line 
for $h\ne 0$ in Fig.~\ref{fig:hPD}. 
This provides a key to resolve the outstanding puzzle about 
the origin of the superconductivity of this material, 
whose transition temperature 
increases even though antiferromagnetic spin fluctuation is suppressed 
under pressure~\cite{kawasaki}.  
Since superconductivity is shown to appear near the VQCP~\cite{OM,WIM}, 
this view verifies a new scenario that the proximity of the VQCP 
is the origin of the superconductivity.

Our result also explains why the metamagneic increase appears in the $m$-$h$ curve 
in $\rm YbAgCu_4$, but not in $\rm YbCdCu_4$, in spite that both have 
the comparable $T_{\rm K}$'s estimated from the $T\to 0$ magnetic susceptibility~\cite{YbInCu4}. 
The relation $h_{\rm m}\sim \Delta T_{\rm K}^{\rm QCP}$ clearly indicates 
that different $h_{\rm m}$ can appear for the same $T_{\rm K}$ 
according to the location in Fig.~\ref{fig:hPD}. 
Namely, the sharp contrast can be understood 
if 
$\rm YbAgCu_4$ and $\rm YbCdCu_4$ are located in the valence-crossover regime 
for $U_{\rm fc}<U_{\rm fc}^{\rm QCP}$ 
and $\Delta T_{\rm K}^{\rm QCP}$ for the former is smaller than that 
for the latter. 
Furthermore, 
the evident FOVT as a function of $T$~\cite{Felner} as well as 
a jump in the $m$-$h$ curve~\cite{metaYbInCu4} 
observed in $\rm YbInCu_4$ is also understood 
if this material is located in the MV regime for 
$U_{\rm fc}>U_{\rm fc}^{\rm QCP}$ in Fig.~\ref{fig:hPD}~\cite{Zlatic}. 
This is consistent with recent band-structure calculations~\cite{harima}. 

\begin{figure}[t]
\includegraphics[width=70mm]{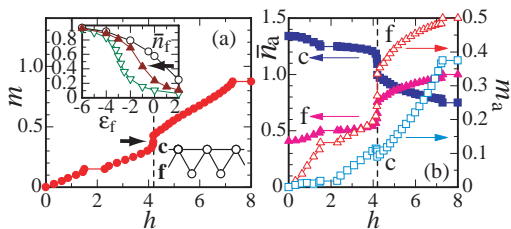}
\caption{\label{fig:DMRG}(color online) 
Magnetization process for $\varepsilon_{k}=-2\cos(k)$,
$V=1$, $U=10^{4}$, $\varepsilon_{\rm f}=-1$ and $U_{\rm fc}=1$ at $n=7/8$:  
(a) $m$-$h$ curve (filled circle). 
An arrow indicates the metamagnetic transition. 
Inset: 
$\varepsilon_{\rm f}$ dependence of $\bar{n}_{\rm f}$ extrapolated to the bulk limit 
for $U_{\rm fc}=0$ (open circle), $U_{\rm fc}=1$ (filled triangle) and $U_{\rm fc}=2$ (open triangle). 
An arrow indicates $\varepsilon_{\rm f}=-1$. 
Lattice structure used in the calculation.
(b) $\bar{n}_{\rm f}$ (filled triangle) and $\bar{n}_{{\rm c}}$ (filled square). 
$m_{\rm f}$ (open triangle) and $m_{\rm c}$ (open square). 
In (a) and (b) dashed lines represent $h=h_{\rm m}$. 
}
\end{figure}

To examine the mechanism more precisely, 
we have applied the density-matrix-renormalization-group method 
to eq.~(\ref{eq:PAM}) in one dimension (1D).  
Since valence fluctuations are basically ascribed to the atomic origin, 
the fundamental properties are expected to be relevant even in 1D~\cite{WIM}. 
We show here the results 
for $\varepsilon_{k}=-2\cos(k)$, $V=1$, $U=10^4$ at $n=7/8$ 
on the lattice with $N=40$ sites for conduction electrons 
illustrated in the inset of Fig.~\ref{fig:DMRG}(a),  
which may be regarded as a 1D mimic of $\rm CeIrIn_5$ and $\rm YbXCu_4$. 
For $h=0$, $\bar{n}_{\rm f}$ vs. $\varepsilon_{\rm f}$ for $U_{\rm fc}=0.0$, 1.0 and 2.0 
is shown in the inset of Fig.~\ref{fig:DMRG}(a). 
As $U_{\rm fc}$ increases, the change in $\bar{n}_{\rm f}$ becomes sharp. 
We show here the magnetization $m$ 
in the MV state of which $\bar{n}_{\rm f}$ at $h=0$ is 
indicated by an arrow in the inset of Fig.~\ref{fig:DMRG}(a). 
A plateau appears at $m=1-n=1/8$, which is expected to disappear 
if we take 
more realistic choice of parameters, e.g. the momentum dependence of 
$V$ and $\varepsilon_{\rm f}$. 
The main result is that metamagnetism emerges as indicated by an arrow. 
The increase of $\bar{n}_{\rm f}$ simultaneous with the decrease of $\bar{n}_{{\rm c}}$ at $h=h_{\rm m}$ 
shown in Fig.~\ref{fig:DMRG}(b) 
reveals that this is caused by the field-induced extension of the QCP to the MV regime. 
Namely, these results indicate that 
the mean-field conclusion is not altered 
even after taking account of quantum fluctuations and electron correlations.  

To explore further the nature of this metamagnetism, 
we have calculated $m_{\rm f}=\sum_{i}\langle S^{{\rm f}z}_i \rangle/N$ and 
$m_{\rm c}=\sum_{i}\langle S^{{\rm c}z}_i \rangle/N$ 
and we find that 
$m_{\rm c}$ decreases at $h=h_{\rm m}$, 
while $m_{\rm f}$ increases as shown in Fig.~\ref{fig:DMRG}(b). 
Since the Kondo cloud is still formed even at $h=h_{\rm m}$, i.e., 
$\langle {\bf S}^{\rm f}_{i}\cdot {\bf S}^{{\rm c}}_{i}\rangle <0$, 
a remarkable decrease of $\langle S^{{\rm c}z}_i \rangle$ is ascribed to 
{\it the field-induced Kondo effect}, which is a consequence of 
the energy benefit by both the Kondo effect and the Zeeman effect. 
Although this mechanism itself has been known to occur in the Kondo regime~\cite{wataKLM}, 
the present finding is that such a mechanism works in the MV regime 
as a driving force of the field-induced VQCP. 

In summary, we have clarified the novel mechanism of the magnetic-field 
dependence of critical points of valence transitions 
and we believe that the proximity of the VQCP revealed in this study provides an essential 
ingredient to understand the $T$-$P$-$h$ phase diagram of the Ce and Yb systems.


\begin{thebibliography}{99}
\bibitem{defMV} We refer to spatially-uniform and quantum-mechanically valence-fluctuating state 
with $\langle n_{\rm f}\rangle<1$ as mixed valence. 
We refer to the state with larger $\langle n_{\rm f}\rangle$ than that 
in the mixed-valence state at the first-order transition as the Kondo state. 
Note that in the intermediate-coupling regime, $\langle n_{\rm f}\rangle$ 
in the Kondo state is smaller than 1.
\end{thebibliography}

\end{document}